\documentclass[11pt,twoside]{article}
\usepackage{FUSE2004}
\usepackage{natbib}
\usepackage{epsfig}
\usepackage{psfig}
\usepackage{lscape}

\newcommand{\FUSE}{{\it FUSE}}
\newcommand{\etal}{et~al.}

\begin{document}
\title{He~II REIONIZATION AND SOURCES OF METAGALACTIC IONIZATION } 
\author{J. Michael Shull}
\affil{CASA, Department of Astrophysical and Planetary
     Sciences, University of Colorado, Boulder, CO 80309
     (mshull@casa.colorado.edu) }

\begin{abstract}
Intergalactic Ly$\alpha$ opacity suggests that H~I was reionized
at $z \approx 6$, while He~II reionization was delayed to $z \approx 3$.  
Both epochs are in conflict with inferences from CMB optical
depth (WMAP) which suggest early reionization at $z = 10-20$.  
One of the major contributions of \FUSE\ to cosmological studies has 
been the detection of He~II Ly$\alpha$ (``Gunn-Peterson") absorption 
in the spectra of AGN at redshifts $z \geq 2.03$.  Spectra of He~II 
absorbers, taken in concert with corresponding H~I (Ly$\alpha$) lines, 
allow us to fix the epoch of helium reionization at 
$z_{\rm HeII} \approx 2.8 \pm 0.2$.  Here, I review \FUSE\ observations 
of He~II absorption, together with their implications for the sources 
and transport of ionizing radiation over 30-50 Mpc distances through 
the IGM.  \FUSE\ observations of He~II absorption toward HE~2347-4342, 
combined with {\it Keck} and {\it VLT} observations of H I,
are consistent with photoionization by QSOs, with a wide range of 
intrinsic spectral indices, and modified by filtering and reprocessing 
in the IGM.  The He~II/H~I ratio ($\eta$) exhibits variations over 1 Mpc 
distance scales ($\Delta z \approx 10^{-3}$).  Intriguingly, this 
$\eta$-ratio is also correlated with Ly$\alpha$ filaments and voids.  
The ionizing radiation field appears to be softer (higher He~II/H~I) 
in the voids.  These void regions may be ionized by local soft sources 
(dwarf starburst galaxies), or the QSO radiation may softened by escape 
from AGN cores and transport through denser regions in the ``cosmic web". 
The differences in ionizing spectra may explain the 1.4 Gyr lag between 
H~I and He~II epochs.

\end{abstract}
\thispagestyle{plain}

\section{Overview: Reionization of the Intergalactic Medium}

Before beginning a discussion of the \FUSE\ contributions
to the study of He~II reionization, I provide an 
overview of the cosmological significance of ``reionization".
This term refers to the sudden increase in ionization fraction
of hydrogen gas in the intergalactic medium (IGM),
initiated at redshifts $z \geq 6$ by the first sources
of ionizing radiation.  After the epoch of recombination  
at $z_{\rm rec} \approx 1300$, the IGM entered a long period 
of ``dark ages" when the IGM was mostly neutral.  This
period ended sometime before $z \approx 6$, perhaps
as early as $z = 10-20$.   

The ionization state of the IGM affects the transmission of
radiation through the Ly$\alpha$ absorbers (H~I and He~II),
as well as the electron-scattering optical depth to the 
cosmic microwave background (CMB) radiation.  As individual
ionizing sources (stars, galaxies, quasars) began emitting 
ionizing radiation, regions of ionized gas were produced,
preceded by ionization fronts (I-fronts) in the IGM.  These
isolated ionized zones (H~II regions, He~III regions) eventually
overlapped (Gnedin 2000), after which the IGM became transparent 
to Ly$\alpha$ radiation.   Residual absorption continued as a 
``picket fence" of discrete, redshifted Ly$\alpha$ absorbers, 
often termed the ``Ly$\alpha$ forest" -- see spectra  
in Songaila (1998). 

The rapid increase in IGM ionization occurred as a topological
phase change in the volume filling factor of ionized zones.  
The longer the first ionizing sources were on, the greater
the volume of the zones. In order to effectively absorb all
the Ly$\alpha$ radiation at $z = 6$ requires a neutral fraction 
of just $x_{\rm HI} = 2 \times 10^{-4}$, even in gas that
is underdense by a factor of 10 relative to the mean.  Simulations 
of this reionization process (Gnedin 2000, 2004) show that the 
transition from neutral to ionized is quite extended in time,
between $z = 5-10$.  The first stage (pre-overlap)  
involves the development and expansion of the first isolated 
ionizing sources. The second stage marks the overlap of the 
I-fronts and the disappearance of the last vestiges of low-density 
neutral gas.  Finally, in the post-overlap stage, the remaining
high-density gas is photoionized from the outside.  The final 
drop in neutral fraction and increase in photon mean free 
path occur quite rapidly, over $\Delta z \approx 0.3$ (see Fig.\ 1 
of Gnedin 2004).   

Reionization is tied closely to the onset of galaxy formation 
beginning at $z \geq 20$.  For the ionization of He~II, and somewhat
for H~I, reionization depends on the rise in the numbers and 
luminosities of active galactic nuclei from $z = 5 \rightarrow 2$.
According to numerical simulations (e.g., Ricotti, Gnedin, \&
Shull 2002, herafter denoted RGS), the earliest epochs of galaxy 
formation began at $z \approx 20-30$ triggered by H~I and H$_2$ 
cooling of protogalactic halos with virial temperatures 
$T_{\rm vir} = 10^3$~K to $10^{4}$~K.  RGS found that the average 
star formation rate density, $\rho_{\rm SFR}$, increased monotonically 
from $10^{-3}$ to $10^{-1}$ $M_{\odot}~{\rm yr}^{-1}~{\rm Mpc}^{-3}$
between $z = 20$ and 10, according to the approximate formula:
\begin{equation}
   {\rm SFR} = (0.003~M_{\odot}~{\rm yr}^{-1}~{\rm Mpc}^{-3})
       10^{(20-z)/5} \; .
\end{equation}

The effects of reionization can be seen indirectly in the 
electron-scattering optical depth of the cosmic microwave background 
(CMB), as we discuss in the next section. However, the most direct 
observations of the IGM ionization state come from their influence on 
the Ly$\alpha$ line-blanketing opacity of the IGM at 
wavelengths shortward of the H~I and He~II Ly$\alpha$ 
lines at $(1215.670~{\rm \AA})(1+z)$ and $(303.782~{\rm \AA})(1+z)$,
respectively. As the amounts of H~I or He~II increase, these 
line-blanketed regions grow into black (``Gunn-Peterson") absorption 
troughs, whose presence can be used to identify high-redshift 
galaxies (Steidel et al.\ 2002).  The best probes of the epoch of 
reionization have come from analyzing the weak transmission of AGN continuum 
flux in these troughs, as the high-redshift IGM transforms from a neutral 
to ionized medium.  The He~II absorption becomes detectable in the far 
ultraviolet at redshifts $z \geq 2.03$ with \FUSE\ and at $z \geq 2.79$ 
with HST.  From the latest Ly$\alpha$ data, the reionization epochs 
occur at $z \approx 6.1\pm0.3$ for H~I (Becker \etal\ 2001; Fan \etal\ 2002;
Gnedin 2004) and at $z \approx 2.8\pm0.2$ for He~II (Kriss \etal\ 2001;
Smette \etal\ 2002; Shull \etal\ 2004; Zheng \etal\ 2004).

\section{Electron-Scattering Optical Depth of the IGM}  

An indirect measure of reionization can be extracted from an analysis
of fluctuations in the CMB radiation, 
particularly the large-angle temperature-polarization correlations.  
By combining the first-year data from the {\it Wilkinson Microwave
Anisotropy Probe} (WMAP) with other results on the CMB, Ly$\alpha$ forest, 
and galaxy large-scale structure, Spergel \etal\ (2003) found a high 
value of the CMB optical depth to electron scattering, 
$\tau_e = 0.17 \pm 0.04$.  This optical depth implies a surprisingly large 
redshift of early IGM reionization, ranging from $11 < z_r < 30$ at 95\% 
confidence (Kogut et al.\ 2003).  More recent estimates of $\tau_e$ and other
cosmological parameters (Tegmark et al.\ 2004) combined the WMAP data 
with the three-dimensional power spectrum from over 200,000 galaxies in 
the Sloan Digital Sky Survey (SDSS).  Their analysis tightened many of the 
WMAP error bars and gave somewhat lower 
$\tau_e = 0.12^{+0.08}_{-0.06}$.  There is still disagreement
between the redshift $z_r \approx 6$ found from (H~I) Ly$\alpha$ opacity
and the ``early reionization" estimates, $z_r =$ 10-20, inferred from the 
CMB.  The optical-depth estimates from WMAP are consistent with 
a wide range in $\tau_e$, reflecting the broad likelihood function 
(see Figure 4 of Spergel \etal\ 2003).  The analysis of second-year WMAP 
data may yield a more precise determination of $\tau_e$.

To elucidate the dependence of $\tau_e$ on the epoch of reionization,
we now provide the following simplified derivation. Assuming instantaneous,
complete ionization at redshift $z_r$, we can write $\tau_e$ as the integral 
of $n_e \sigma_T d \ell$, the electron density times the Thomson cross section 
along proper length.  For a standard $\Lambda$CDM cosmology, we have 
$(d \ell /dz) = c(dt/dz) = (1+z)^{-1} [c/H(z)]$, where 
$H(z) = H_0 [\Omega_m (1+z)^3 + \Omega_{\Lambda}]^{1/2}$ and
$\Omega_m + \Omega_{\Lambda} = 1$.  We adopt
$n_e = n_H (1+y)$ and $n_H = [(1-Y) \rho_{\rm cr}/m_H](1+z)^3$,
where $Y = 0.244$ (He mass fraction) and $y \approx 0.0807$ 
(He fraction by number). We further assume that helium is singly ionized 
(He~II at $z \geq 3$).  The above integral can be done exactly: 
\begin{eqnarray}
   \tau_e &=& \left( \frac {c}{H_0} \right) \left( \frac {2 \Omega_b}
      {3 \Omega_m} \right)  
     \left[ \frac {\rho_{\rm cr} (1-Y)(1+y) \sigma_T } {m_H} \right] 
     \left[ \{ \Omega_m (1+z_r)^3 + \Omega_{\Lambda} \}^{1/2} 
      - 1 \right]   \nonumber \\ 
  & \approx & (0.0376 h) \left( \frac {\Omega_b} {\Omega_m} \right)
      \left[ \{ \Omega_m (1+z_r)^3 + \Omega_{\Lambda} \}^{1/2} - 1 
        \right]  \; . 
\end{eqnarray} 
In the above numerical expression, we can substitute the WMAP-concordance
values (Spergel et al.\ 2003; Bennett \etal\ 2003), 
$\Omega_b h^2 = 0.0224 \pm 0.0009$, $\Omega_m h^2 = 0.135^{+0.008}_{-0.009}$, 
and $h = 0.71^{+0.04}_{-0.03}$, where 
$h = [H_0/(100~{\rm km~s}^{-1}~{\rm Mpc}^{-1}$] is the scaled
Hubble constant.  The critical density is
$\rho_{\rm cr} = (1.879 \times 10^{-29}$~g~cm$^{-3}) h^2$. 
For large redshifts, $\Omega_m (1+z)^3 \gg \Omega_{\Lambda}$, 
and the integral simplifies to 
\begin{equation}
  \tau_e \approx \left( \frac {c}{H_0} \right) \left( \frac {2 \Omega_b}
   {3 \Omega_m} \right)  \left[ \frac {\rho_{\rm cr} (1-Y)(1+y) \sigma_T} 
   {m_H} \right] \Omega_m^{1/2} (1+z_r)^{3/2} \approx (0.00229) (1+z_r)^{3/2} \; ,  
\end{equation}  
nearly independent of $h$.  Additional contributions to $\tau_e$ arise from 
residual electrons ($10^{-3.3}$ ionized fraction) frozen out after recombination
(Seager, Sasselov, \& Scott 2000) and from partial reionization prior 
to $z_r$ by the first stars (Venkatesan, Tumlinson, \& Shull 2003) and early
X-ray sources (Venkatesan, Giroux, \& Shull 2001; Ricotti \& Ostriker 2004).  
These additional ionization sources could contribute extra electron
scattering in the amount $\Delta \tau_e \approx 0.02$.   
We now invert the above relation to estimate the reionization redshift,
\begin{equation}
   (1+z_r) \approx (16.2) \left( \frac {\tau_e}{0.15} \right)^{2/3} \; .  
\end{equation} 
Here, we scaled to $\tau_e = 0.15$, equal to the WMAP-measured value of 
optical depth, $\tau_e = 0.17 \pm 0.04$, reduced by $\Delta \tau_e = 0.02$ 
arising from additional electron scattering prior to $z_r$.  A re-analysis 
of WMAP data (Tegmark \etal\ 2004), including prior knowledge of the SDSS 
galaxy power-spectrum, found slightly lower optical depth, 
$\tau_e = 0.124^{+0.083}_{-0.057}$ and reionization at 
$z_r = 14.4^{+5.2}_{-4.7}$.  The large ($1 \sigma$) likelihood range 
allows an optical depth near the value, $\tau_e \approx 0.05$, 
expected from from eq. (4), assuming a reionization epoch 
$z_r \approx 6-7$ consistent with H~I Ly$\alpha$ opacity.  The agreement 
would be particularly good if one allows for sources of partial ionization 
at $z > z_r$.

\section{FUSE Observations of He~II Absorbers}

Over the last decade, IGM studies have been enhanced by high-resolution 
optical and UV spectra of quasar absorption lines.  
Studies in H~I and He~II Ly$\alpha$ are particularly important for 
understanding the transition of the 
high-redshift IGM from a neutral to ionized medium.  The He~II 
absorption becomes detectable in the far ultraviolet at redshifts 
$z \geq 2.03$ with \FUSE\ and at $z \geq 2.79$ with the
{\it Hubble Space Telescope} (HST).  Moderate-resolution \FUSE\ spectra 
of He~II absorption (Kriss \etal\ 2001; Shull \etal\ 2004; Zheng \etal\ 
2004) confirm the general theoretical picture (Cen \etal\ 1994; Dav\'e 
\etal\ 2001) of the IGM as a fluctuating distribution of baryons, 
organized by the gravitational forces of dark matter and photoionized by
high-redshift quasars and starburst galaxies (Haardt \& Madau 1996;
Fardal, Giroux, \& Shull 1998, hereafter FGS).

The He~II reionization epoch appears to occur at $z \approx 2.8\pm0.2$
(Kriss \etal\ 2001; Smette \etal\ 2002; Shull \etal\ 2004),   
based on several sightlines studied with HST and \FUSE.  This effect
is demonstrated in Figure 1, which shows the 
patchy regions of broad-band (He~II Ly$\alpha$) opacity 
between $z = 2.7-2.9$ toward the QSO HE~2347-4342.
Helium reionization is delayed by 1.4 Gyr ($\Delta z \approx 3$)
from hydrogen reionization, probably because He~II is
more difficult to ionize, requiring a harder (4 ryd continuum),
probably from AGN.  It is also possible that partial He~II reionization 
could occur at $z > 6$, from a population of ultra-hot, zero-metallicity
massive stars (Tumlinson \& Shull 2000;
Venkatesan, Tumlinson, \& Shull 2003).

\begin{figure}[h]
\plotone{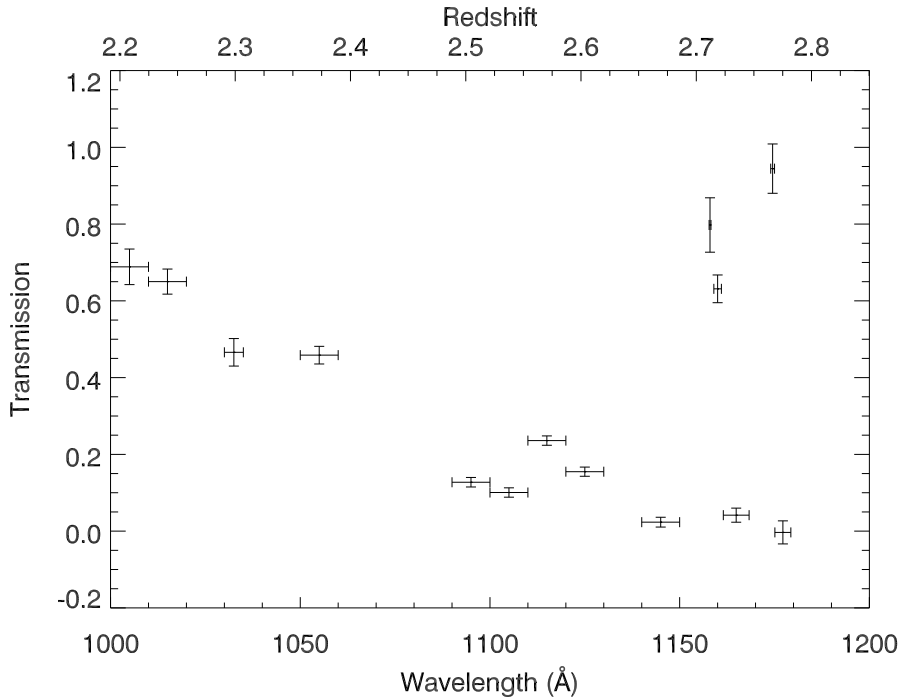}
\caption{\small The epoch of gradual He~II reionization at 
$z_{\rm HeII} \approx 2.8 \pm 0.2$ toward HE~2347-4342 (Shull \etal\
2004) is illustrated by the continuum flux transmission 
through He~II absorption in several well-defined redshift windows.  Note 
the three regions of high transmission between $z = 2.7-2.8$, 
flanked by low transmission regions. The transmission slowly 
recovers at $z < 2.6$.  }
\end{figure}

\FUSE\ observations of He~II can be
combined with the corresponding H~I Ly$\alpha$ lines at optical 
wavelengths, to yield the ratio, $\eta$ = N(He~II)/N(H~I). As we explain, 
measurements of $\eta$ probe the shape of the high-redshift metagalactic 
radiation field, presumably from quasars and massive stars.  
Figure 2 shows a portion of the combined observations of He~II and H~I
absorption for the Ly$\alpha$ forest at $z \approx 2.3-2.6$.   
The He~II absorption is generally much stronger than H~I by a factor 
$\eta = $N(He~II)/N(H~I), predicted theoretically (FGS) to be 50--100
for photoionization by quasar backgrounds.  For optically thin lines,
we can write $\eta \approx 4 \tau_{\rm HeII}/\tau_{\rm HI}$.
The greater strength of He~II
arises because it is harder to photoionize than H~I, owing to lower fluxes
and cross sections at its ionizing threshold ($h\nu_T = 54.4$~eV)
and because He~III recombines 5 times faster than H~II.
Variations in $\eta$ can be used to diagnose the sources
and fluctuations of the metagalactic ionizing spectrum in the
continua of H~I (1~ryd) and He~II (4~ryd).
He~II absorption is also a good diagnostic of absorption
from low-density regions (``voids") in the IGM.  Because 
$\tau_{\rm HeII} = (\eta/4) \tau_{\rm HI}$ ($\eta \gg 1$), 
He~II can be tracked into much lower density regions than H~I.
He~II absorption can also be used to probe collisionally
ionized gas at $T \approx 10^5$~K, produced by shock-heating
during structure formation (Cen \& Ostriker 1999;  Dav\'e et al.\ 2001).

\begin{figure}[h]
\plotone{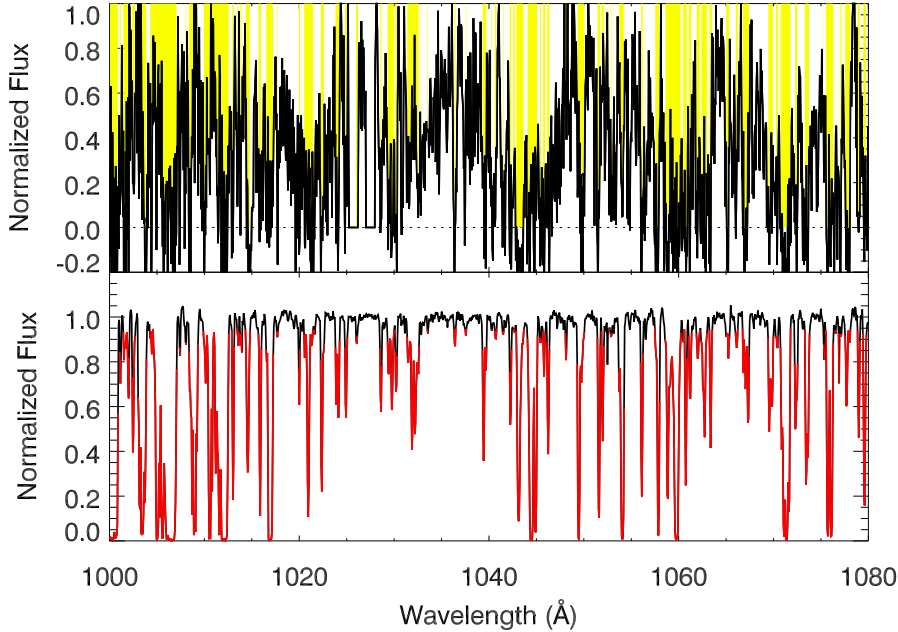}
\caption{\small Overlay of the Ly$\alpha$ absorption of He~II (top panel 
-- \FUSE) and of H~I (bottom panel -- VLT/UVES) toward HE~2347-4342.  
In the VLT data, ``filament" points are defined as those with 
$\tau_{\rm HI} > 0.05$.  Wavelengths of H~I Ly$\alpha$ were divided
by 4 for alignment. Note the strong He~II absorption (top). } 
\end{figure}

Following the initial \FUSE\ discovery and analysis paper (Kriss \etal\ 2001),
Shull \etal\ (2004) and Zheng \etal\ (2004) performed in-depth analyses of 
the He~II and H~I absorption toward HE 2347-4342, an AGN at redshift 
$z_{\rm em} = 2.89$. We probed the IGM at redshifts $z$ = 2.3--2.9, 
using spectra from \FUSE\ and the {\it Ultraviolet-Visual Echelle Spectrograph} 
(UVES) on the VLT telescope. We focussed on two major topics:
(1) small-scale variability ($\Delta z \approx 10^{-3}$) in the ratio
$\eta =$ N(He~II)/N(H~I); and (2) an observed correlation of high-$\eta$
absorbers (soft radiation fields) with voids in the (H~I) Ly$\alpha$
distribution.  These $\eta$ variations (Figure 3) probably reflect fluctuations 
in the ionizing sources on scales of 1 Mpc, modified by radiative transfer 
through a filamentary IGM whose opacity controls the penetration 
of 1-5 ryd radiation over 30--40 Mpc distances.
In photoionization equilibrium, the He~II/H~I ratio can be expressed:
\begin{equation}
   \eta = \frac {n_{\rm HeIII}} {n_{\rm HII}}
          \frac {\alpha_{\rm HeII}^{(A)}} {\alpha_{\rm HI}^{(A)}}
          \frac {\Gamma_{\rm HI}} {\Gamma_{\rm HeII}}
   \approx (1.70) \frac {J_{\rm HI}} {J_{\rm HeII}}
        \frac {(3 + \alpha_4)}{(3 + \alpha_1)} T_{4.3}^{0.055} \; .
\end{equation}
Here, $\alpha_{\rm HI}^{(A)}$, $\alpha_{\rm HeII}^{(A)}$,
$\Gamma_{\rm HI}$, and $\Gamma_{\rm HeII}$ are the case-A recombination
rate coefficients and photoionization rates for H~I and He~II, and
$J_{\rm HI}$ and $J_{\rm HeII}$ are the specific intensities
of the radiation field at the H~I (912~\AA) and He~II (228~\AA) edges.
The parameters $\alpha_1$ and $\alpha_4$ are the local spectral indices
of the ionizing background at 1 and 4 ryd, respectively, which
provide minor corrections to the photoionization rates.
We adopt case-A hydrogenic recombination rates to H~I and He~II,
appropriate for the Ly$\alpha$ forest absorbers with very low neutral
fractions. Over the temperature range $16,000~{\rm K} < T < 32,000$~K,
we approximate the case-A recombination rates coefficients as
$\alpha_{\rm HeII} = (1.36 \times 10^{-12}~{\rm cm}^3~{\rm s}^{-1})
T_{4.3}^{-0.700}$ and
$\alpha_{\rm HI} = (2.51 \times 10^{-13}~{\rm cm}^3~{\rm s}^{-1})
T_{4.3}^{-0.755}$, where $T_{4.3} \equiv (T/10^{4.3}~{\rm K})$.
We assume that $n_{\rm He}/n_{\rm H} = 0.08$ ($Y = 0.244$) and
that H and He are fully ionized ($n_e = 1.16 n_H$)
after reionization ($z < 3)$.

\begin{figure}[h]
\plotone{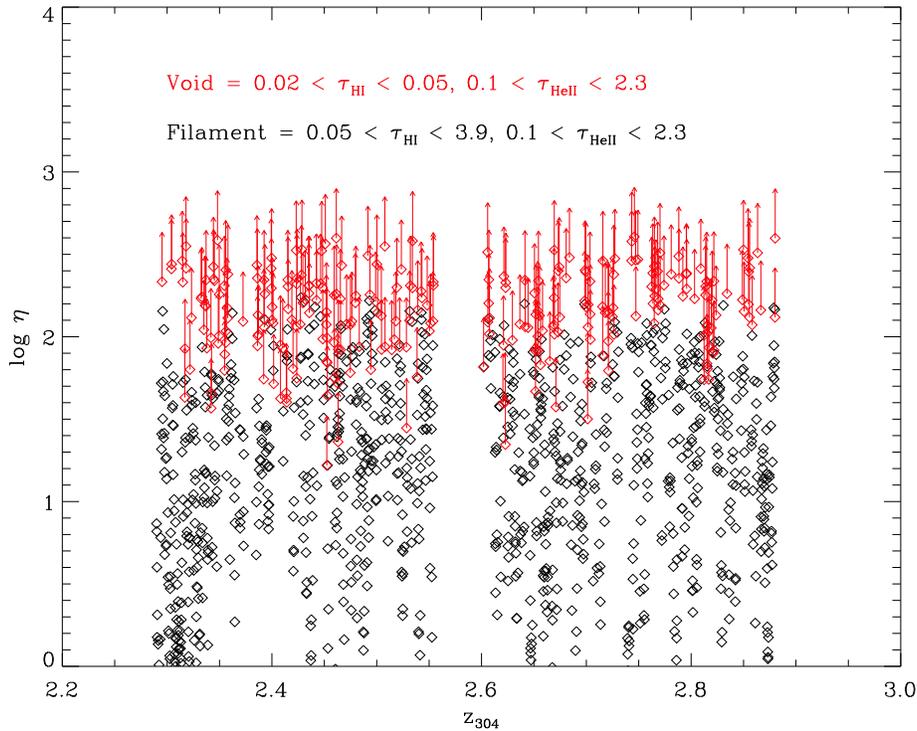}
\caption{\small Observed wide variations in the ratio 
$\eta$ = N(He~II)/N(H~I) in the IGM toward HE~2347-4342 
(Shull \etal\ 2004).  These variations imply substantial 
fine-grained fluctuations in the ionizing radiation field at 
1 ryd (H~I) and 4 ryd (He~II) on scales of $\Delta z = 0.001$ or 1 Mpc.}
\end{figure}

Owing to photon statistics and backgrounds, we can measure optical depths
over the ranges $0.1 < \tau_{\rm HeII} < 2.3$ and 
$0.02 < \tau_{\rm HI} < 3.9$, and we can reliably determine values of 
$\eta \approx 4 \tau_{\rm HeII}/\tau_{\rm HI}$ over the range 0.1 to 460.  
Values $\eta = 20-200$ are consistent with
models of photoionization by quasars with observed spectral indices
ranging from $\alpha_s = 0-3$ (Figure 4).  
Values $\eta > 200$ may require additional contributions
from starburst galaxies, heavily filtered quasar radiation, or density variations.
Regions with $\eta < 30$ may indicate the presence of local hard sources.
We find that $\eta$ is higher in ``void'' regions, where H~I is weak or
undetected and $\sim$80\% of the path length has $\eta > 100$. These voids
may be ionized by local soft sources (dwarf starbursts) or by QSO radiation
softened by escape from the AGN cores or transfer through the ``cosmic web".
The apparent differences in ionizing spectra may help to explain the
1.4 Gyr lag between the reionization epochs of H~I
($z_{\rm HI} \sim 6.1 \pm 0.3$; Gnedin 2004) and He~II 
($z_{\rm HeII} \sim 2.8 \pm 0.2$; Kriss \etal\ 2001; Shull \etal\ 2004).

\begin{figure}[h]
\plotone{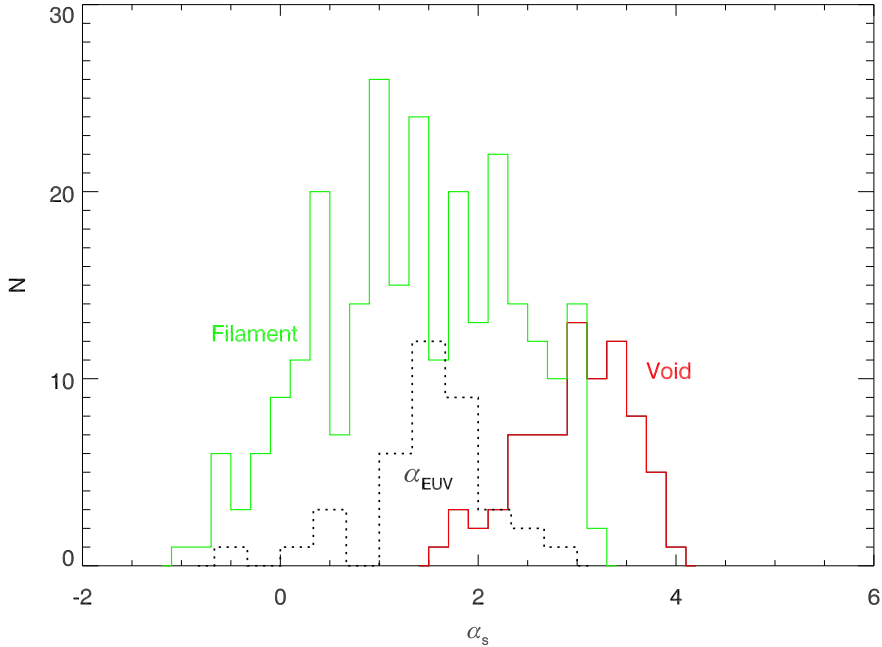}
\caption{\small Observed distribution of the observed QSO ionizing spectral
indices, $\alpha_{\rm EUV}$, for radio-quiet AGN observed by HST 
(Telfer et al.\ 2002).  Other curves show the spectral indices $\alpha_s$,
with $J_{\nu} \propto \nu^{-\alpha_s}$, that are required
to reproduce the observed values of He~II/H~I ratios, using eq. [5]. 
Note the systematically softer radiation fields (high $\eta$, large 
$\alpha_s$) in the voids ($\tau_{\rm HI} < 0.05$) compared to filaments
($\tau_{\rm HI} > 0.05$). The breadth of these distributions and the
clear offset between voids and filaments suggest that additional spectral
softening may be present. } 
\end{figure}

Many of the variations in the ionizing radiation field appear to
come from intrinsic differences in QSO spectra (Fig.\ 4),
found by direct measurements of the redshifted EUV continua using HST 
(Telfer \etal\ 2002) and \FUSE\ (Scott \etal\ 2004).  
Figure 5 shows some of the actual AGN spectra on which these
conclusions are based.  It is apparent that even low-redshift
QSOs and Seyferts have wide differences in their EUV continuum slopes,
which may be luminosity dependent (Scott \etal\ 2004).  
In addition to intrinsic source variations, the metagalactic radiation 
field is heavily filtered and reprocessed by intervening IGM (see FGS).  
Because the He~II opacity is so strong ($\eta \gg 1$), fluctuations at
4 ryd are probably larger than those at 1 ryd.  Unfortunately, we must
conclude that the sources, spectra, and fine-grained spatial variations 
in the metagalactic ionizing radiation field are still not well understood.  

Future studies of He~II should proceed along several fronts.
First and foremost, we need to find additional high-redshift AGN targets 
for which we can repeat the He~II/H~I measurements and $\eta$ analysis.  
Quasar H1700+6416 at $z_{\rm em} = 2.74$ is one such \FUSE\ target, but 
more are needed to identify the cosmic variance in $z_r$(He~II).  
Additional theoretical work is needed, using cosmological
simulations that include radiative transfer of the 1 ryd and 4 ryd 
radiation intensities and their spatial fluctuations.  Radiation
in the 3-4 ryd band may be especially important for the
photoionization corrections to C~III, C~IV, Si~III, and Si~IV 
on which IGM metallicities are based.  Finally, it would be helpful 
to characterize the mean intrinsic EUV spectra of AGN, at $z \leq 0.5$, 
which are still highly uncertain (Figure 5) despite heroic efforts with
both HST and \FUSE.  There is currently only a small wavelength window 
at which the rest-frame EUV is observable, and it does not extend to 
$\lambda \leq 228$ \AA, the 4 ryd continuum that controls He~II ionization.

\begin{figure}[h]
\plotone{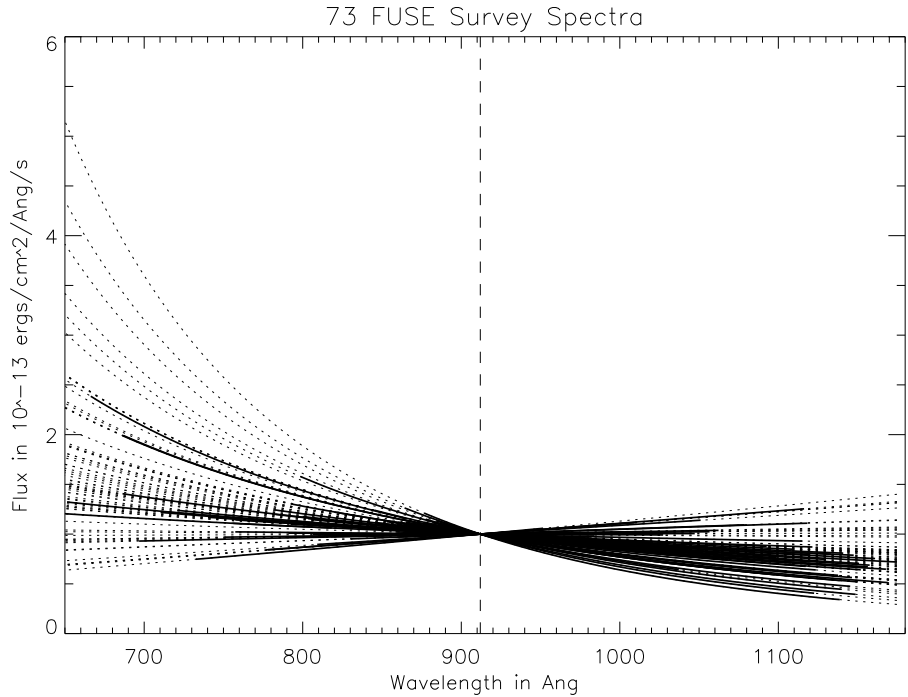}
\caption{Normalized distribution of fits to the rest-frame UV and EUV 
continuum spectra of 73 AGN (Clausen \& Shull 2004), taken from \FUSE\ data 
analyzed and fitted by Scott \etal\ (2004).  We have shifted these data to 
the rest-frame, K-corrected the UV fluxes, and normalized them to unity 
at 1 ryd (912 \AA).  The solid lines show actual data, and dotted lines
are extrapolations.  Note the wide variations in spectral indices
at $\lambda = 650-900$~\AA. }   
\end{figure}

\end{document}